\documentclass[aps, twocolumn,floatfix,showpacs, superscriptaddress]{revtex4-2}

\usepackage{amssymb,amsmath,amsfonts}
\usepackage{graphicx,graphics}
\usepackage{epsfig}
\usepackage{epstopdf}
\usepackage{amsfonts}
\usepackage{bm,bbm}
\usepackage{amssymb,amsmath,amsfonts}
\usepackage{mathrsfs}
\usepackage{calc}
\usepackage{epsfig}
\usepackage{float}
\usepackage{xcolor}
\usepackage{epstopdf}
\usepackage{bm,amssymb,amsmath}
\usepackage{graphicx,color}

\usepackage{ulem}
\usepackage{xspace}

\newcommand{\ie}{information engines }

\newcommand{\IE}{Information Engines }
\newcommand{\se}{Szil\'ard engine }
\newcommand{\s}{Szil\'ard }
\begin{document}

\title{Experimental Realizations of Information Engines: Beyond Proof of Concept}
\author{R\'emi Goerlich}
\affiliation{Raymond \& Beverly Sackler School of Physics and Astronomy, Tel Aviv University, Tel Aviv 6997801, Israel}
\author{Laura Hoek}
\affiliation{Raymond \& Beverly Sackler School of Chemistry, Tel Aviv University, Tel Aviv 6997801, Israel}
\author{Omer Chor}
\affiliation{Department of Physics of Complex Systems, Weizmann Institute of Science, Rehovot 7610001, Israel}
\author{Saar Rahav}
\affiliation{Schulich Faculty of Chemistry, Technion--Israel Institute of Technology, Haifa 3200008, Israel}
\author{Yael Roichman}
\email{roichman@tauex.tau.ac.il}
\affiliation{Raymond \& Beverly Sackler School of Physics and Astronomy, Tel Aviv University, Tel Aviv 6997801, Israel}
\affiliation{Raymond \& Beverly Sackler School of Chemistry, Tel Aviv University, Tel Aviv 6997801, Israel}

\date{\today}
\begin{abstract}
    Gathering information about a system enables greater control over it. This principle lies at the core of information engines, which use measurement-based feedback to rectify thermal noise and convert information into work. Originating from Maxwell's and Szilárd's thought experiments, the thermodynamics of information engines has steadily advanced, with recent experimental realizations both confirming established results and pushing the field forward. Coupled with technological advances and developments in nonequilibrium thermodynamics, novel implementations of information engines continue to challenge theoretical understanding. In this perspective, we discuss recent progress and highlight new opportunities, such as applying information engines to active, many-body, and inertial systems, and leveraging tools like optimal control to design their driving protocols.
\end{abstract}

\maketitle

\section{Introduction}

Information is now widely acknowledged as a measurable physical quantity with thermodynamic implications. This notion was initially introduced in a thought experiment designed by James C. Maxwell in 1867 \cite{maxwell1871theory}. Maxwell proposed that if it is possible to observe an isolated system, gather information about its state, and subsequently take actions based on the acquired information, it allows for the reduction of the total entropy of the system, seemingly defying the Second Law of Thermodynamics.
Taking it a step further, Leo Szil\'ard, in 1929, proposed a refined design that directly converts information gained through measurement into extracted energy, creating an engine that operates using a single heat bath \cite{szilard1929entropieverminderung, szilard1964decrease}. Thus, Szil\'ard's \textit{information engine} appears to challenge the fundamental principle governing heat engines, which states that they can operate only between two heat reservoirs at different temperatures.

From Maxwell and Szil\'ard to its later refined versions, the various models of \ie can be understood by information thermodynamics and, more generally, non-equilibrium statistical mechanics \cite{Parrondo_2015, lutz_information_2015, paneru_colloidal_2020}.
Within this unifying framework, information forms a thermodynamic resource. When information is gathered, it drives the system out of equilibrium, creating a state from which energy can be extracted. This led to an extension of the Second Law of Thermodynamics to account for the role of information in processes involving measurement and feedback \cite{sagawa_generalized_2010, esposito_second_2011, ashida_general_2014}. Nowadays, the operational cycle of information engines applied to thermal systems is well understood from a thermodynamic perspective \cite{lutz_information_2015}.

\begin{figure}
    \centering
    \includegraphics[scale = 0.17]{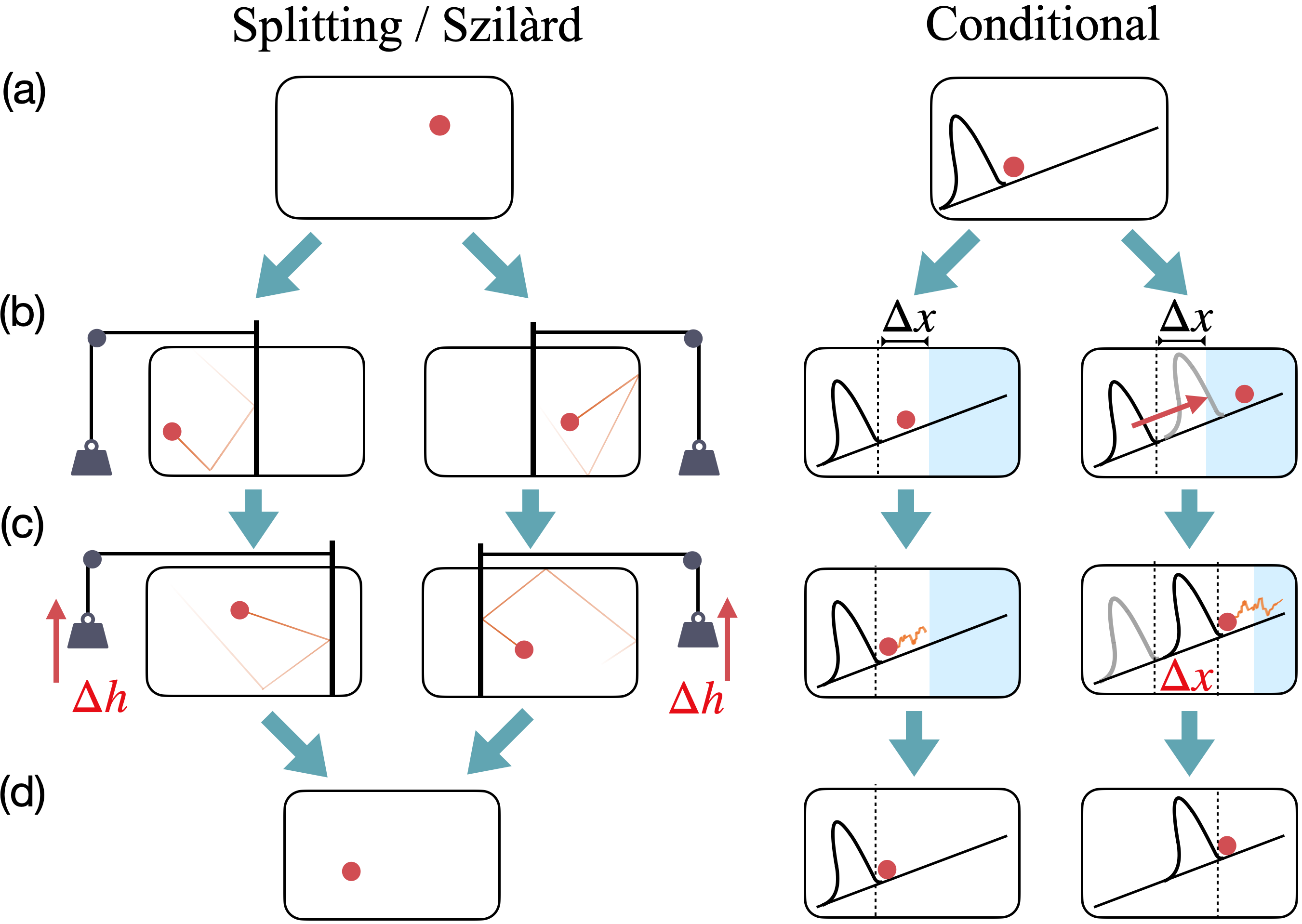}
    \caption{The operational sequence of a phase space splitting (on the left) and a conditional, phase-space contracting (on the right) information engine is as follows: The cyclic operation begins at some initial state of the system (a). The system's micro-state is measured, and suitable feedback is applied (b). Subsequently, the system is allowed to relax (c), followed by the erasure of the measurement outcome, returning the system to a different but equivalent initial state (a).}
    \label{fig:IEs_sketch}
\end{figure}

More precisely, the action sequence of \ie can vary in its details, but it always involves the following steps depicted in Fig.~\ref{fig:IEs_sketch}: (a) measuring the state of a system, (b) making a decision based on the measurement as to which feedback should be applied to it, if any. (c) Allowing the system to respond to the applied feedback and, in doing so, extracting energy. Finally, if one wants to bring both the engine and the measurement apparatus back to their initial state, (d) erasing the measurement outcome \footnote{In many experimental realizations \cite{sagawa_generalized_2010, admon_experimental_2018, saha2021maximizing}, the measurement-feedback scheme allows to store potential energy, which is not necessarily extracted, which also releases the constrain of erasing the memory.}.

The increasing experimental control over fluctuating systems enabled the physical realization of information engines.
Such engines were realized on a microscopic scale using colloidal suspensions \cite{toyabe_experimental_2010, roldan_universal_2014, paneru_lossless_2018}, nanoscale solid-state structures \cite{koski_experimental_2014, chida_power_2017}, and DNA \cite{Ribezzi_Crivellari_2019}. On the macroscopic scale, granular gas \cite{lagoin_human-scale_2022} and self-propelled bristle robots \cite{Chor2023} were used.
Early experimental studies of information engines primarily sought to demonstrate the fundamental relationship between information and thermodynamics. Subsequent realizations focused on enhancing the output power or the efficiency of information engines, leveraging increasingly precise control mechanisms \cite{ashida_general_2014, paneru_optimal_2018, lucero_maximal_2021, saha2021maximizing, saha_bayesian_2022}. More recently, experiments with information engines have aimed to push the field in new directions.
This arise from the parallel development of stochastic thermodynamics on heat engines \cite{martinez2017colloidal}, active matter \cite{ramaswamy2003active, fodor2021active}, fluctuation theorems \cite{seifert2012stochastic, ciliberto2017experiments}, and optimal control \cite{schmiedl2007optimal, rosales2020optimal, pires2023optimal}. Emerging experimental concerns have also posed new theoretical challenges, such as addressing finite-size effects, particularly critical in systems involving active matter \cite{elgeti2015physics, krishnamurthy2016micrometre, solon2015pressure}, and managing finite-time operations, which demand the development of new thermodynamic optimization principles \cite{ParkOptimal2016, davis2024active}.

These observations motivate this perspective, in which we review the most salient aspects of experimental realizations of information engines and highlight the current evolution of this field [Sec.~\ref{Sec:Experimental}].
We emphasize new directions related to non-equilibrium working substance, many-body effects, inertia, and optimal control with finite resources [Sec.~\ref{Sec:NewHorizons}].
While various processes and phenomena can be cast into the information engine formalism \footnote{This includes Brownian ratchets \cite{reimann2002brownian}, autonomous systems, diode-like systems \cite{raizen_comprehensive_2009} or even apparently unrelated stochastic processes \cite{goerlich2023experimental, fuchs2016stochastic}. All are able to rectify fluctuations and can therefore be described using the \ie formalism. The current perspective keeps a more restricted scope.}, this perspective centers on \ie with explicit experimental realizations of feedback loops \cite{toyabe_experimental_2010, roldan_universal_2014,  paneru_lossless_2018, lee_experimentally-achieved_2018, Ribezzi_Crivellari_2019, paneru_efficiency_2020, saha2021maximizing, saha2023information, archambault2024inertial, archambault2024first}.

\section{Modern \IE}
\label{Sec:Experimental}

\subsection{Two Classes of Information Engines}

Szil\'ard's engine is a paradigmatic information engine in which each measurement splits the phase-space into two parts, and feedback is systematically applied. Its operation sequence is depicted in Fig.~\ref{fig:IEs_sketch}, on the left. The engine is comprised of an ideal gas molecule in a chamber in contact with a heat reservoir (Fig.~\ref{fig:IEs_sketch}a, left). After each measurement, a partition is inserted into the chamber, splitting it into two (Fig.~\ref{fig:IEs_sketch}b, left). A measurement is used to determine in which half the molecule resides, thereby updating the state of the system. A mass is connected to the partition, on the side populated by the molecule. The gas is then allowed to expand isothermally, performing work by lifting the mass (Fig.~\ref{fig:IEs_sketch}c, left). Finally, the mass is detached from the partition, the measurement outcome is erased, and the system returns to its original state (Fig.~\ref{fig:IEs_sketch}d, left).
Engines of the Szil\'ard type, namely phase-splitting \ie have been realized using various microscopic systems \cite{koski_experimental_2014, roldan_universal_2014, Ribezzi_Crivellari_2019}, and recently also macroscopic systems \cite{Chor2023}.

In a second class of \ie a condition set on the measurement outcome determines whether feedback should be applied.
More precisely, feedback is applied only when it is beneficial for the extraction of energy.
An example of a conditional information engine is shown in Fig.~\ref{fig:IEs_sketch} (right). A repelling optical force (a wall) prevents a diffusing particle from going down a linear potential. If, when measured, the particle is further up than a given threshold distance $\Delta x$ from the repelling wall (Fig.~\ref{fig:IEs_sketch}b (rightmost panel)), the wall is moved closer to the particle, confining it into a smaller region of phase space. Thus, the average position of the particle increases against the potential's slope, without applying direct work. Conditional \ie have been realized in various optical tweezers set-ups \cite{admon_experimental_2018, paneru_lossless_2018, lee_experimentally-achieved_2018, saha2021maximizing, saha2023information} including the first realization \cite{toyabe_experimental_2010}.

Szil\'ard's engines convert all measurement outcomes into mechanical work while conditional \ie are storing potential energy thanks to selected measurement outcomes. 
Such classification is open to debate \footnote{in terms of phase-space manipulation, compressing phase-space into a smaller region can be viewed as an extreme case of Szil\'ard-like splitting, both schemes sharing some similarities. More crucial is the fact that measurements are selected in the case of conditional engine} yet we argue that categorizing them separately reveals crucial insights. This distinction hinges on two key factors: the type of phase-space manipulation and the nature of the feedback timing, leading to very different maximal work extraction per measurement.

\subsection{Thermodynamics of \IE}
\label{Sec:Thermo}

The Second Law of Thermodynamics bounds energy exchanges.
The average work $\Delta W$ needed to perform an isothermal transformation between equilibrium states $\rho_{\rm eq}$ is bounded by free energy
\begin{equation}
    \Delta W \geq \Delta F = \Delta E - T \Delta S,
    \label{eq:EqSecondLaw}
\end{equation}
where $\Delta E$ is the average energy difference between initial and final states and $\Delta S$ the associated entropy difference $S$, defined as $S(\rho) = -k_B \int \rho \ln \rho ~dx$ \cite{Seifert2005}.
We introduced here the Second Law on averaged quantities, yet work, heat, and entropy are all well defined at the level of single fluctuating trajectories, by stochastic thermodynamics \cite{Sekimoto1998, SekimotoBook, seifert2012stochastic}.
Eq.~(\ref{eq:EqSecondLaw}) is the consequence of Jarzynski equality relating the fluctuations of the stochastic work $\Delta w$ to the free-energy as $\langle e^{-(\Delta w-\Delta F)/k_B T} \rangle = 1$ \cite{Jarzynski1997}.
The Second Law also constrains the operation of information engines, but it must be refined with respect to Eq~(\ref{eq:EqSecondLaw}) to describe the use of information properly.
In a seminal work, Sagawa and Ueda derived a Second Law-like inequality for information engines \cite{sagawa_generalized_2010}.
Following the operation steps of a Szil\'ard engine clearly illustrates this generalization.

\begin{figure}
    \centering
    \includegraphics[scale = 0.24]{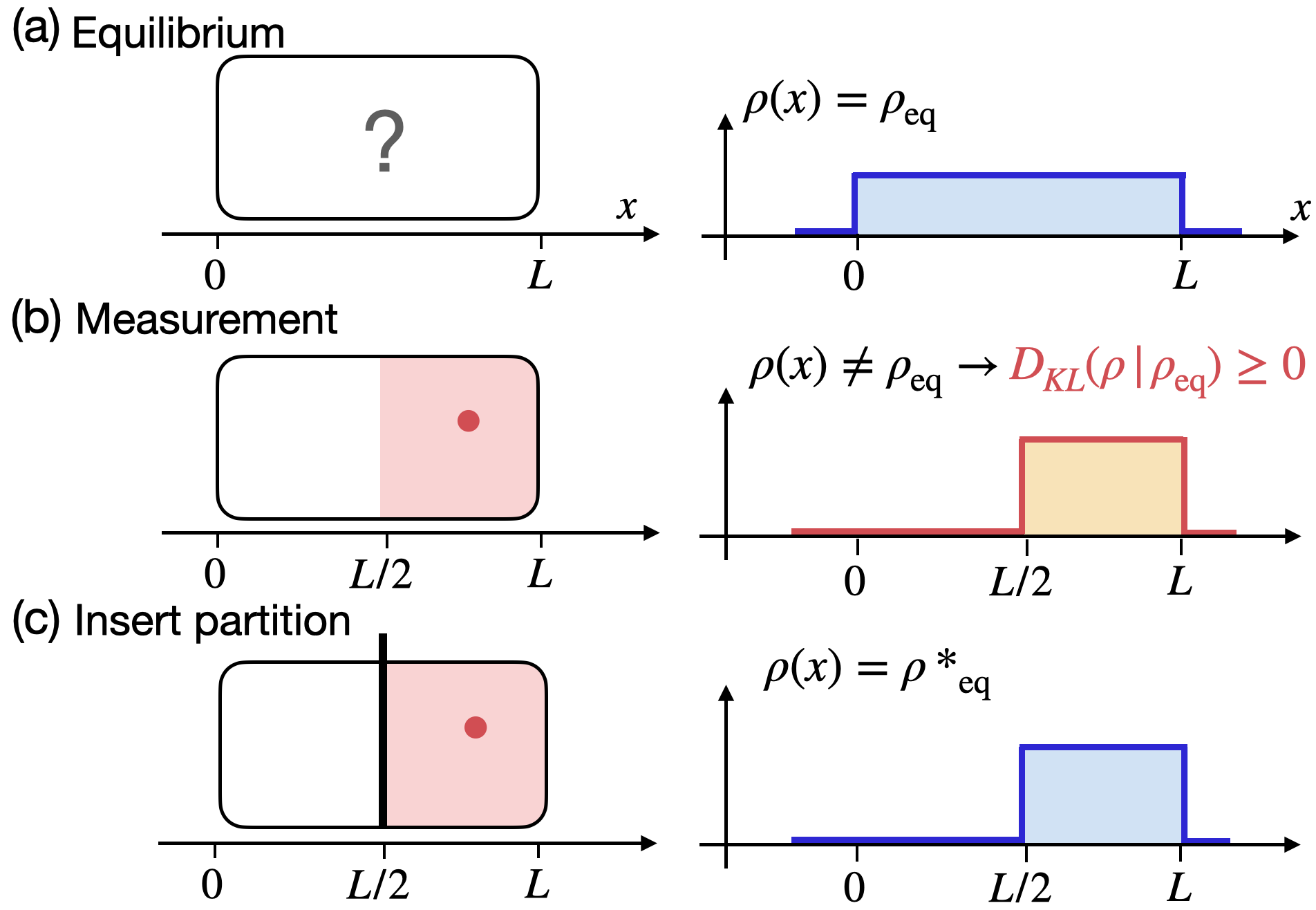}
    \caption{The first steps of operation of a \se from a thermodynamic point of view. (a) initially at equilibrium, the position of the particle is unknown. Its probability density $\rho(x)$ (blue line) spans uniformly the whole available volume. As such, it corresponds to the equilibrium Boltzmann distribution $\rho_{\rm eq}$. (b) a binary measurement collapses $\rho(x)$ in one half of the box. The post-measurement distribution now differs from the equilibrium solution, revealed by a positive information.
    (c) Inserting a partition reduced the available volume, and the post-measurement $\rho(x)$ corresponds to equilibrium again \cite{Parrondo_2015}. The engine is then quasistatically expanded, and the whole operation remains reversible, extracting $W^{\rm out}_{\rm rev} = k_B T \ln(2)$.}
    \label{fig:Szilard}
\end{figure}
Initially at thermal equilibrium (see Fig.~\ref{fig:Szilard} (a)), the probability distribution of the particle's position $\rho_{\rm eq}$ is uniform in the box.
After applying a measurement with a binary outcome (left or right), the probability distribution $\rho$ becomes non-zero only in one half of the box.
$\rho$ is therefore a non-equilibrium state with lower entropy, obtained by acquiring information about the system (Fig.~\ref{fig:Szilard} (b), right panel).
The generalization of the Second Law Eq.~(\ref{eq:EqSecondLaw}) accounting for a non-equilibrium initial state reads \cite{esposito_second_2011}
\begin{equation}
    \Delta W \geq \Delta F - k_B T I.
    \label{eq:NeqSecondLaw}
\end{equation}
$I$ quantifies the information acquired by the measurements as $I = \sum_{m} p(m) \mathcal{D}_{\rm KL}(\rho(x|m) | \rho_{\rm eq})$ where $p(m)$ is the probability of a measurement outcome $m$ (such as the particle being measured in a certain region of the box).
$\mathcal{D}_{\rm KL}(\rho(x|m) | \rho_{\rm eq}) \equiv \int \rho(x|m) \ln [\rho(x|m) / \rho_{\rm eq}(x)] dx \geq 0$ is the Kullback-Leibler (KL) divergence between the non-equilibrium state $\rho(x|m)$ after measurement and the equilibrium state $\rho_{\rm eq}(x)$ before measurement.
$I$ is therefore the average KL-divergence over the possible outcomes. (see \footnote{The Second Law generalized to transformations between non-equilibrium states reads in general $\Delta W \geq \Delta F + k_B T \Delta I$ where $\Delta F$ and $\Delta I$ corresponds to the difference in equilibrium free energy and KL-divergence between the final and initial states \cite{esposito_second_2011}. For \ie the measurement induces an initial non-equilibrium state, while the feedback action usually returns the system to equilibrium. In that case $\Delta I = I(t_f) - I(t_i) = -I(t_i)$ and we obtain $\Delta W \geq \Delta F - k_B T I(t_i)$ as is used in the main text in most of the literature.} for a note on sign convention in $I$).
As such, the thermodynamics of \ie is no more than a special case of non-equilibrium thermodynamics \cite{Parrondo_2015}.
For a \se as shown Fig.~\ref{fig:Szilard}, the information gathered by a binary measurement reads $I = \ln(2)$ and this initial distance to equilibrium is the thermodynamic resource from which work can be extracted \footnote{For a Szil\'ard engine operating on a box of length $L = 1$ we have explicitly $\rho_{\rm eq}(x) = 1$ on $x \in [0, 1]$. The measurement can have two outcomes: either left (l) with $\rho(x|l) = 2$ on $x \in [0, \frac{1}{2}]$ or right (r) with $\rho(x|r) = 2$ on $x \in [\frac{1}{2}, 1]$. The information reads
$I = \sum_m p(m) D_{\rm KL}(\rho(x|m)||\rho_{\rm eq}(x))dx$. In the case where both outcomes are equiprobable $p(l) = p(r) = 1/2$ one gets  $I = \frac{1}{2} \int_0^{1/2} 2 \ln(2) dx + \frac{1}{2} \int_{1/2}^2 2 \ln(2) dx = \ln(2)$ and hence $W^{\rm out} = k_B T \ln(2)$.}.
If the measurements suffer from errors, $I$ is replaced by the mutual information between the system and the measurement device \cite{KoskiMutual2024}.

The remaining steps of an ideal \se consists of reversibly transforming the energy contained in the non-equilibrium state into mechanical work, ending up back in the initial state.
To do so, (Fig.~\ref{fig:Szilard} (c)) a partition is inserted in the box, and the state $\rho$ instantaneously becomes an equilibrium state again in the new, smaller volume.
It can then be quasistatically driven back to the initial state, extracting work and closing the operation cycle.
As this cycle leaves the free energy unchanged $\Delta F = 0$, the maximal energy produced reads $ \Delta W^{\rm out}_{\rm rev} = k_B T I$ where $W^{\rm out} = - W$ is the work extracted from the system.
Here, it corresponds to $W^{\rm out}_{\rm rev} = k_B T \ln(2)$.

At the level of stochastic trajectories, the Sagawa-Ueda equality generalizes Jarzinsky equality in the presence of information as $\langle e^{-(\Delta w-\Delta F)/k_B T - j} \rangle = 1$ where $j$ is the stochastic information, with $I = \langle j \rangle$ \cite{Sagawa2010}.
While we used Szil\'ard's engine to illustrate Eq.~(\ref{eq:NeqSecondLaw}) on Fig.~\ref{fig:Szilard}, this principle extends beyond information engines to encompass all transformations between non-equilibrium states \cite{esposito_second_2011}. The positive Kullback-Leibler divergence $\mathcal{D}_{\rm KL}$ measuring the deviation from equilibrium enables these systems to exceed the performance limits of classical heat engines. This example demonstrates that work can be extracted even when the engine operates within a single heat bath.

\subsection{Experimental Realizations of Information Engines}
\label{Sec:History}

\begin{figure} [h!]
    \centering
    \includegraphics[scale = 0.235]{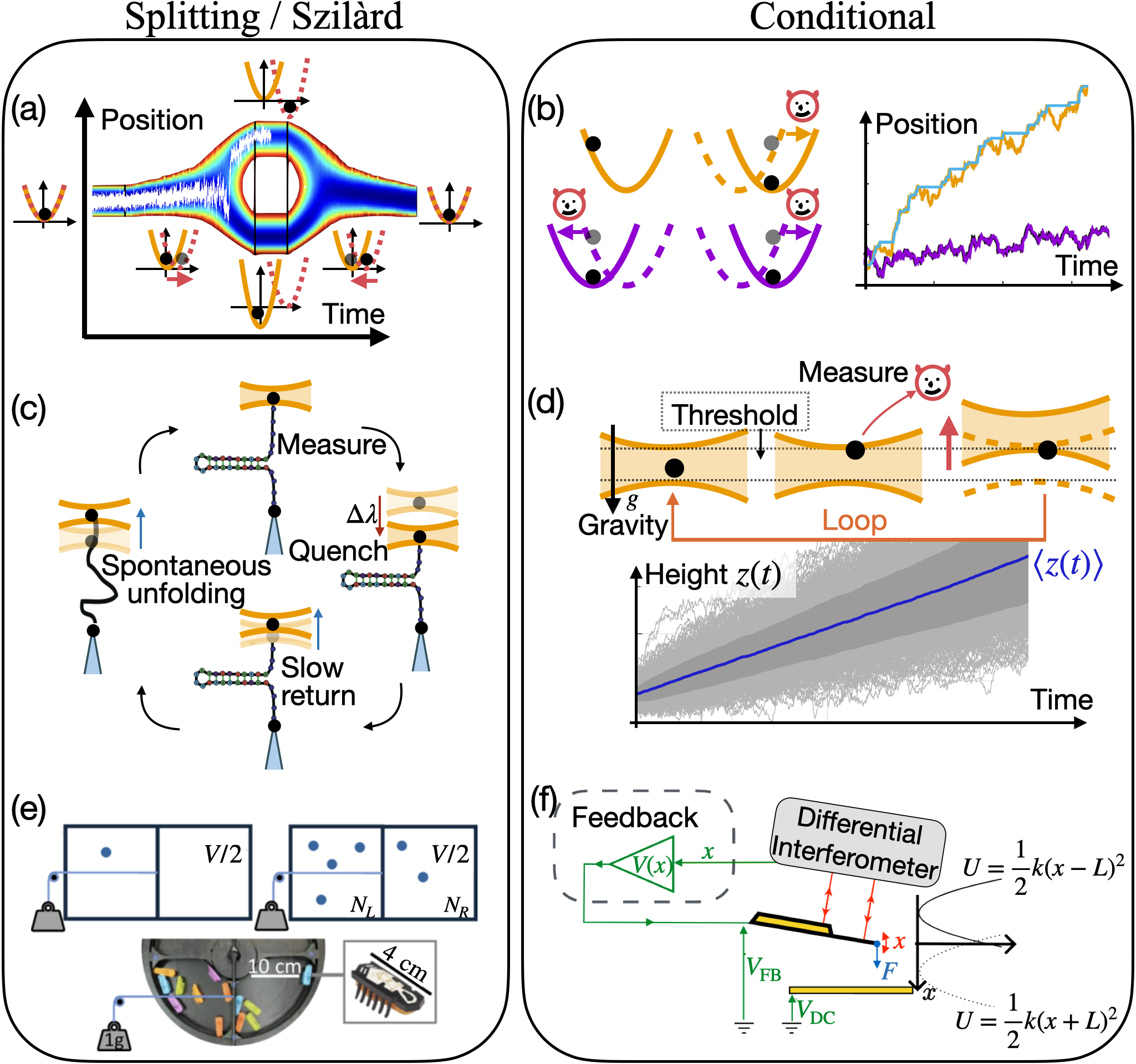}
    \caption{Experimental realizations of information engines, sorted into splitting (a,c,e) and conditional (b, d, f) categories.
    (a) A colloidal particle is manipulated with a static and  mobile optical trap. After measuring the position of the particle the potential landscape is tilted accordingly, quenching the state of the system. Adapted from \cite{roldan_universal_2014}.
    (b) If the traped colloidal particle is measured above a defined threshold, the potential is shifted by the same amount. Transport is obtained only if the condition is applied in one direction (yellow lines), otherwise it produces cooling of the particle in the potential's co-moving frame (purple lines). Adapted from \cite{lee_experimentally-achieved_2018}.
    (c) A DNA hairpin is held in an optical tweezers. The logic of operation is similar to the first example above, but bi-stability is a result of the spontaneous folding and unfolding of the DNA, instead of a couple of optical traps. Adapted from \cite{Ribezzi_Crivellari_2019}.
    (d) A heavy colloidal particle is held against gravity, when it is measured above a threshold, the trap is shifted upward by twice this amount, ensuring zero work exchange.  Adapted from \cite{saha2021maximizing}.
    (e) Many-body active Szil\'ard engine \cite{Chor2023}, based on macroscopic artificial active matter composed of bristle-bots. The number of bots in left and right halves of the box is measured, and feedback is applied by placing a partition in the middle of the box and connecting a weight to it so it would be lifted when the partition is pushed towards the less dense half.
    (f) Conditional engine in the inertial regime \cite{archambault2024first}. The fluctuations of a micron-scale cantilever in vacuum are monitored and a feedback potential allows to switch the center of the effective harmonic potential, condition by the crossing of a defined threshold by the cantilever.}
    \label{fig:Experiments}
\end{figure}

The operation of an information engine requires real-time measurements of the system's fluctuations and the capability to apply feedback based on these measurements. spans various length scales, from microscopic to macroscopic, enabling the experimental realization of information engines of both the phase space splitting type (Fig.~\ref{fig:Experiments} (a,c,e)) and the conditional information engine type (Fig.~\ref{fig:Experiments} (b,d,f)) across different scales. Early implementations of Maxwell's demon \cite{serreli_molecular_2007, raizen_comprehensive_2009}, led the way to the first experimental realization of an information engine \cite{toyabe_experimental_2010}, which converted information into potential energy.

Explicit \ie operating on the nanometer scale have been realized with two very different experimental systems.
On the one hand, using a pair of nanometric metallic islands, forming a single electron box \cite{koski_experimental_2014}.
In this work, one extra electron can fluctuate between both metallic islands, its state is monitored by a single-electron transistor electrometer and the gate voltage applied to the islands is modified accordingly.
In doing so, the electron is captured in its measured state, before being slowly released, extracting energy from its thermal hopping between islands.
On the other hand, using a stretched DNA hairpin (Fig.~\ref{fig:Experiments}~(c)) \cite{Ribezzi_Crivellari_2019}, rectifying the fluctuations between folded and unfolded states, monitored and manipulated using optical tweezers. Both realizations use a Szil\'ard-like phase-space splitting protocol and share the following logic: the system fluctuates between two metastable states. Upon measurement, the system is quenched in its current state by a strong external potential. Work is extracted from thermal jumps during the quasistatic return to the original bistable state.

The majority of realized information engines operate on the mesoscopic micrometer scale, where the working substance is a single colloidal particle in a fluid. Here, Brownian motion rectification is performed by optical tweezers and electric fields. One of the early realizations was based on a similar phase-space splitting protocol as the examples above, where bistability is induced through a double-well optical potential (Fig.~\ref{fig:Experiments}~~(a)) \cite{roldan_universal_2014}. In this setup, the distance between both wells and their depths is adjusted to quench spontaneous fluctuations and extract work. Several other mesoscopic information engines were designed to convert information into translation, entering our second category: conditional feedback, compressing phase space. These engines achieved energetically cost-less transport (Fig.~\ref{fig:Experiments}~(b)) \cite{paneru_lossless_2018, lee_experimentally-achieved_2018}, climbing up a potential landscape (Fig.~\ref{fig:Experiments}~(d)) \cite{toyabe_experimental_2010, saha2021maximizing}, and even translation against a fluid current \cite{admon_experimental_2018}. In these engines, the operating mode involves monitoring the particle's position within the effective potential at regular time intervals. When a beneficial fluctuation is detected, such as the particle moving in the desired direction, feedback is applied by adjusting the potential to prevent the particle from reverting to its previous state.

Both types of information engines, phase-space splitting, and conditional feedback, were realized on the macroscopic scale as well. Notably, fluctuations at this scale are not thermal but instead result from external or internal driving forces, making the working substance (system and heat reservoir) far from thermal equilibrium. At the macroscopic scale, the environmental degrees of freedom become accessible, allowing for quantification and manipulation of the bath itself. 

A macroscopic conditional information engine was realized using a blade in a driven granular gas as the working substance \cite{lagoin_human-scale_2022}, which is typically characterized by thermal-like velocity distribution \cite{puglisi2012structure, gnoli_brownian_2013}. In this realization, collisions between the granular gas and an immersed blade were rectified electrically to result in an angular drift of the blade. A Szil\'ard's-like splitting engine was realized using bristle robots (bbots) as a many-body working substance (Fig.~\ref{fig:Experiments}~(e)) \cite{Chor2023}. bbots are a class of active matter that exhibits correlated motion \cite{deblais2018boundaries, giomi2013swarming} resulting in anomalously large density fluctuations \cite{ramaswamy2003active, ginelli2010large}. In this experiment, the number of particles on each side of a chamber was measured. A partition was then inserted in the middle of the chamber, and a mass was connected to it in the direction of the most populated side. Work was extracted by the motion of the partition towards the less populated half of the system, which raised the mass against gravity.

\subsection{Key Aspects of Operation: From Symmetry to Efficiency}

Spatial and temporal symmetries are central in the operation of information engines.
For example, at the beginning of the cycle of a \s engine, as depicted on the left panel of Fig.~\ref{fig:IEs_sketch} (a) and (b), the left/right measurement corresponds to a spatial symmetry breaking of the probed phase-space \cite{horowitz2013optimizing}.
Conversely, the end of the same cycle (Fig.~\ref{fig:IEs_sketch} (c) and (d)) corresponds to the associated symmetry restoration.
The change in entropy during the breaking and recovery of spatial symmetry dictates the thermodynamics of the engine \cite{horowitz2013optimizing, roldan_universal_2014}.
Interestingly, this original \s engine, which operates through spatial symmetry breaking, can achieve 100\% efficiency in converting information to work \cite{szilard1929entropieverminderung,Parrondo_2015}.
This is the consequence of another symmetry aspect: the temporal symmetry of its ideal cycle of operation, i.e., its thermodynamic reversibility \cite{horowitz2010nonequilibrium, horowitz2013optimizing, ashida_general_2014}.

In contrast with the ideal operation of Szil\'ard engine, many information engines lack reversibility, leading to a lower value for the maximal work extracted: they do not reach equality in the Second Law  Eq.~(\ref{eq:NeqSecondLaw}).
Reversibility can break due to numerous effects, some of them intrinsically connected to the operation of the engine.
This motivated the definition of a modified version of the Second Law, explicitly incorporating irreversibility
\begin{equation}
    \Delta W \geq \Delta F - k_B T (I - I_u)
    \label{eq:GenNeqSecondLaw}
\end{equation}
where $I_u$ is called \textit{unavailable} information \cite{ashida_general_2014}.
It quantifies irreversibility as the relative fraction of time-reversed paths that do not exist in the forward operation of the feedback.
More precisely, a phase-splitting Szil\'ard engine (Fig.~\ref{fig:IEs_sketch} (left)), follows the steps (1) to (4) as detailed above.
Its backward operation corresponds to reinserting the partition from the side of the box and slowly compress the single-particle gas back to half the full volume.
It ends up in the initial post-measurement state, showing the reversibility of the engine's operation.
In most experimental realizations however, the box is formed by a potential, typically quadratic \cite{roldan_universal_2014, toyabe_experimental_2010, Ribezzi_Crivellari_2019, archambault2024inertial}.
In that case, when reinserting the partition from a finite position, there will always be a non-zero probability that the particle lies beyond it \cite{archambault2024inertial}.
In that \textit{singular} case, compressing the volume again does not lead to the initial post-measurement state, since the particle lies on the wrong side of the partition \cite{Horowitz_2010, horowitz2011thermodynamic}.
Eq.~(\ref{eq:GenNeqSecondLaw}) accounts for the associated decrease of available work.
Notably, these singular paths arise by construction for any conditional phase-contracting information engines (Fig.~\ref{fig:IEs_sketch} (right)).
Time-reversal symmetry is also broken for many-body \se  \cite{horowitz2013optimizing, Chor2023} which will be presented in the final section of this perspective.

Finally, due to the feedback mechanism, the system's dynamic is correlated with the measurements.
It is therefore affected by measurement errors such as detection uncertainties and the temporal delay in the feedback.
This is another source of irreversibility progressively degrading the work output, eventually reaching a threshold where work extraction becomes impossible \cite{saha_bayesian_2022}. This limitation has been overcome using inference techniques: by implementing a Bayesian approach that leverages past measurements to better predict current system states, researchers have achieved robust power output even in noisy conditions \cite{saha_bayesian_2022}.

The generalized Second Law Eq.~(\ref{eq:GenNeqSecondLaw}) is derived from a generalized Jarzynski equality \cite{ashida_general_2014}.
It provides a tight upper bound on the maximum work extractable by irreversible information engines, as verified experimentally in refs.~\cite{paneru_lossless_2018, archambault2024first}.

Reversely, natural correlations and memory in the dynamics can be used to increase the output power of conditional engines, by adapting the rate of measurement.
This was explored for a microsphere under the influence of gravity in ref.~\cite{saha2021maximizing} (Fig.~\ref{fig:Experiments}d).
Measuring the particle's position very frequently eventually allows for the rectification of any positive fluctuation in the system.
Similarly, continuously monitoring the movement of a DNA hairpin and its correlations allows the rectification of rare crossing events, increasing the power extraction \cite{Ribezzi_Crivellari_2019}.
However, information processing has an energetic cost, as formalized by Landauer's principle \cite{Parrondo_2015}.
This is the motivation for the definition of the information-to-work conversion efficiency of an \ie
\begin{equation}
    \eta = \frac{W^{\rm out}}{k_B T (I-I_u) + \sum_i W_i}
\end{equation}
relating the useful information $I-I_u$ and the output work $W^{\rm out}$.  $\sum_i W_i$ is a shorthand for all additional costs of operating the engine: dissipation during non-quasistatic transformations, or spurious work applied on the system.
Just like regular thermal engines \cite{schmiedl2007efficiency}, \ie can be designed to maximize a trade-off between power and efficiency.
This was explored for a particle transported across space \cite{lee_experimentally-achieved_2018} and against a fluid flow \cite{admon_experimental_2018}.
In both experiments, it is demonstrated that an optimal rate of measurement exists that allows for the mitigation of information costs and the increase in power extraction.

Dynamic correlations become particularly significant in active self-propelled systems and underdamped dynamics. In these cases, the relationship between engine operation and power extraction follows more intricate patterns \cite{archambault2024inertial, archambault2024first, saha2021maximizing}. We will examine these complexities in detail in the final section of this perspective.

\subsection{Different Approaches for Work Extraction}

In experimental Szil\'ard-like \ie, work is extracted from \ie in the return to the engine's initial state (Fig.~\ref{fig:IEs_sketch}c).
In the experiment mentioned above \cite{koski_experimental_2014, Ribezzi_Crivellari_2019,roldan_universal_2014}, work is extracted when the system slowly returns to a symmetric equilibrium state. However, the extracted energy is not coupled to an external system but rather dissipated into heat.

In experimental conditional information engines, two cases must be distinguished.
First, in transport on a flat energy surface \cite{lee_experimentally-achieved_2018}, no work is extracted, but rather, the natural cost of the operation is avoided.
Second, if the transport is exerted against some potential, energy is effectively stored \cite{toyabe_experimental_2010, admon_experimental_2018, saha2021maximizing}.
The stored energy could be extracted by letting the particle flow down the potential but here again this energy would be dissipated as heat if the particle is not coupled to an external device.

Only a few experimental realizations coupled the engine to an external system to harvest mechanical work. Examples include the macroscopic Brownian ratchet \cite{lagoin_human-scale_2022} where a blade immersed in granular gas was coupled to a motor to create a current, as well as a macroscopic many-body \se \cite{Chor2023}, where weight was lifted against gravity.
Notably, these two experiments are macroscopic systems, based on athermal fluctuations.
In the latter \cite{Chor2023}, controlling the load in time further allows to approach the quasistatic limit, where work extraction is maximal.
Constructing energy transduction at the microscopic scale is possible but more challenging due to the fluctuating nature of devices in this scale, aligning with Feynman's historical refutation of the ratchet and pawl mechanism \cite{feynman1965feynman, bang2018experimental}.

\section{New horizons}
\label{Sec:NewHorizons}

Experimental realization of information engines are tightly connected to the parallel development of stochastic thermodynamics, progressively incorporating more complex and rich working substances.
New challenges have emerge, but also innovative approaches to enhance engine performance. Three key research directions have shaped the field: \textbf{A.} non-thermal or active working substances and many-body systems, \textbf{B.} inertial effects, and \textbf{C.} optimal control.

\subsection{Active Matter and Collective Effects in \IE}

Models of active particles, such as bacterial swarms or animal flocking, were introduced to mimic biological phenomena \cite{vicsek1995novel, solon2024thirty}. These systems, composed of energy-converting agents like bacteria and artificial self-propelled particles, are now a key research area in stochastic thermodynamics \cite{ramaswamy2010mechanics, fodor2018statistical, fodor2016far, fodor2021active, Albay2O23, davis2024active}

A minimal model for active particles involves a particle interacting with correlated and potentially non-Gaussian noise. Such non-equilibrium fluctuations can arise naturally from internal mechanisms, such as bacterial self-propulsion \cite{di_leonardo_bacterial_2010}, or be incorporated in a controlled manner by external driving forces \cite{Goerlich2022, saha2023information}. The presence of correlated noise alone can generate superdiffusion and increase effective temperatures, enhancing work extraction and efficiency in heat engines \cite{holubec2020active, krishnamurthy2016micrometre, martinez_colloidal_2017}.
Consequently, several theoretical studies predict a substantial enhancement in the performance of \ie when active self-propelled particles are considered \cite{paneru_optimal_2018, paneru_colossal_2022, malgaretti_szilard_2022, cocconi_efficiency_2024, rafeek2024active, garcia2024optimal}.

The performance of \ie ultimately relies on the temperature difference between the hot working substance and the cold measurement-feedback device, which enables precise tasks. One of the key effects of non-equilibrium driving noise is the large increase in the system's effective temperature due to large fluctuations \cite{Goerlich2022, di_leonardo_bacterial_2010}. These excess fluctuations can be harvested to extract energy, similar to biological molecular motors that harness large non-equilibrium fluctuations \cite{ariga2021noise}. As discussed in \cite{saha2023information}, rectifying non-equilibrium fluctuations can also be scaled to macroscopic systems, such as turbulent flows \cite{francois2020nonequilibrium}, potentially opening new avenues for energy harvesting.

Beyond non-equilibrium fluctuations, the emergence of complex collective behaviors, driven by the properties of the constituents, such as shape, number, propulsion mechanisms, and couplings, gives rise to exotic phases such as flocking \cite{deseigne2010collective, liebchen2017collective, liebchen2018synthetic}, mobility-induced phase separation \cite{cates2015motility}, swarming \cite{aranson2022bacterial, giomi2013swarming}, boundary effects \cite{deblais2018boundaries}, giant number fluctuations \cite{ramaswamy2003active, ginelli2010large}, and non-reciprocal interactions \cite{loos2020irreversibility, dinelli2023non, fruchart2021non}. 
These phenomena will significantly influence the operation of information engines, offering numerous possibilities for design and control. The development of novel information engines can harness two key phenomena: the long temporal persistence induced by activity and the large spatial correlations in collective effects. This approach brings information engines closer to biological systems, where information processing naturally occurs within an active environment.

Activity also interplays non-trivially with many-body effects in information engines. Consider again the example of the Szil\'ard engine: what happens when the number of particles in the chamber increases? Can more work be extracted? Experimental evidence shows that the answer depends significantly on whether the particles are ideal gas molecules or active self-propelled particles \cite{Chor2023}. For ideal gas molecules, the probability of finding them on either side of the box follows a binomial distribution. As the number of molecules increases, relatively small imbalances in the density, that result in even smaller work extraction, become more likely. Consequently, the average work extracted per measurement decreases as the number of molecules grows.
In stark contrast, active particles exhibit giant number fluctuations. As a result, when measuring the state of a many-body active Szil\'ard engine, there is a high probability of finding an uneven distribution of particles, leading to significant work extraction. In this case, the average work per measurement increases with the number of active particles \cite{Chor2023}.

\subsection{Inertial Effects}

Another significant advancement in stochastic thermodynamics is its application to underdamped stochastic systems, such as levitating particles in optical tweezers \cite{rondin2017direct, gonzalez2021levitodynamics, piotrowski2023simultaneous} and micro-cantilevers \cite{dago2021information, dago2022dynamics, dago2023adiabatic, barros2024probabilistic}. These experimental platforms extended the range of accessible thermal phenomena \cite{rademacher2022nonequilibrium, Dechant2015, rondin2017direct}, opening new avenues towards  the thermodynamics of feedback \cite{debiossac2020thermodynamics} and of quantum systems,  via ground state cooling \cite{Brunelli2018}. On a macroscopic scale, inertial effects natuarally arise in granular gases \cite{puglisi1999kinetic, puglisi2002fluctuation}.

Underdamped \ie have been explored recently using micro-cantilevers (Fig.~\ref{fig:Experiments}~(f)) \cite{archambault2024inertial}.
In this conditional information engine, a micro-cantilever fluctuates in an external potential and its position is measured at regular intervals of duration $\tau$. If the measured value exceeds a set threshold, the potential is shifted in the same direction, similarly to the protocol in Fig.~\ref{fig:Experiments}~(b) \cite{paneru_lossless_2018, lee_experimentally-achieved_2018}. This protocol enables two distinct operations.
First, if the potential is shifted by exactly twice the measured position, the potential energy is conserved, but the system will relax around a shifted equilibrium position. This leads to net transport with zero work injection.
Second, if the potential is shifted by a smaller amount, the transport is reduced, but a net work is extracted at each application of the feedback.

Significant differences between overdamped and underdamped systems arise for finite $\tau$, which is the condition for finite power output. In the overdamped case, the power monotonically decreases as $\tau$ increases \cite{park2016optimal} while in the underdamped case, power is maximal for a finite $\tau$.
Resonances with relaxation time can cause a collapse in power output.
In the case where potential energy (e.g. gravity) is stored at each step \cite{saha2021maximizing} the interplay between parameters described in \cite{archambault2024inertial} could be leveraged to further maximize energy harvesting.
With a measurable velocity, inertial systems also provide new routes to the optimal design of \ie \cite{saha2021maximizing}.

\subsection{Optimal Control}

To extract finite power from \ie (or any thermodynamic machine), finite-time operation is required. However, energy dissipation during a thermodynamic transformation only vanishes in the case of a slow reversible process, which saturates the Second Law, Eq.~(\ref{eq:NeqSecondLaw}). 
Optimal control theory aims to identify protocols that reduce the time required for a transformation while minimizing dissipation \cite{schmiedl2007optimal, gomez2008optimal, rosales2020optimal, pires2023optimal, guery2023driving, loos2024universal, baldovin2024optimal}. Applying optimal control to the operation of \ie should prove crucial, as it will enable rapid operation with minimal losses, thereby increasing both power output and efficiency.

In the operation of an information engine, both the timing of measurements and the protocol of work extraction can be controlled to optimize the operation of the engine.

The most straightforward parameter governing the performance of an \ie is the measurement rate, which tunes the trade-off between power and efficiency \cite{saha2021maximizing, admon_experimental_2018}.
The decrease in efficiency at high rates is rooted in the cost of information erasure \cite{landauer1961irreversibility, lutz_information_2015}.
While most studies of \ie used the theoretical minimal Landauer cost for erasure in efficiency assessments, practical \ie rarely achieve this ideal limit. Several experiments focused specifically on the issue of minimizing dissipation in erasure, both in the quasistatic regime \cite{berut_experimental_2012, klaers_landauers_2019} and more recently, in finite-time \cite{dago2023adiabatic}.

The second aspect of optimization considers the feedback and work extraction protocols, usually via time manipulation of the potential.
Specifically, seeking to optimize state-to-state transformation from the non-equilibrium state just after measurement to the final equilibrium state, which closes the engine's cycle.
New optimization procedures are needed to design such protocols minimizing dissipation over finite times \cite{esposito_second_2011}.
This task is even more challenging for active working substances \cite{davis2024active}.

A few initial theoretical works have focused on optimizing the operation of information engines.
In ref.~\cite{kamijima2024finite}, the information flows between the system and the demon is manipulated to minimize dissipation.
In ref.~\cite{garcia2024optimal} an optimal protocol to control active particles is derived, and used to build a cyclic information engine, which efficiency depends on the specific parameters of the active particle involved.
In general, the optimal operation of information engines depends on system-specific parameters. Interestingly, the relevant parameters can be inferred from the engine's measurements, enabling a dynamic approach where the operation is continuously adjusted based on real-time data.
The use of reinforcement learning \cite{Whitelam2023, barros2024learning} further extends the range of application, beyond analytical solutions.
It could lead to the development of model-free optimal information engines using the result of past measurement to optimize the future operation.

\section{Conclusions}
\label{Sec:Conclusions}

In this perspective, we discussed recent advancements and emerging directions for information engines, emphasizing experimental realizations across scales from nanometric to macroscopic. We reviewed their working principles, their thermodynamics, and the role of reversibility in optimizing work extraction, focusing on their experimental realization. Key future directions are underlined.
First, the integration of active matter and mesoscopic many-body systems in \ie  enhance power extraction through non-equilibrium noise and complex collective phenomena. Additionally, inertial effects induce intricate relationships between operational and relaxation times, demanding refined measurement protocols. Finally,optimal control theory and machine learning represent new paths to enhance the performance and adaptively optimize information engines.

\section*{Acknowledgments}
We sincerely thank Cyriaque Genet for insightful discussions.
Y.R. acknowledges support from the Israel Science Foundation (grants No. 385/21). Y.R, L.H, and R.G. acknowledge support from the European Research Council (ERC) under the European Union’s Horizon 2020 research and innovation programme (Grant agreement No. 101002392). R.G. acknowledges support from the Mark Ratner Institute for Single Molecule Chemistry at Tel Aviv University.
SR is grateful for support from the the Israel Science Foundation (Grant No. 1929/21).


%

\end{document}